\documentclass[lettersize,journal]{IEEEtran}
\usepackage{amsmath,amssymb,amsfonts}
\usepackage{array}
\usepackage{textcomp}
\usepackage{stfloats}
\usepackage{url}
\usepackage{verbatim}
\usepackage{graphicx}
\usepackage{cite}
\usepackage{xcolor}
\def\BibTeX{{\rm B\kern-.05em{\sc i\kern-.025em b}\kern-.08em
    T\kern-.1667em\lower.7ex\hbox{E}\kern-.125emX}}

\usepackage{siunitx}  
\graphicspath{{figure/}}
\usepackage{float}
\usepackage{subcaption}
\usepackage{soul}
\usepackage{xcolor}

\usepackage{breakurl}
\usepackage[breaklinks]{hyperref}

\usepackage{amsthm}
\usepackage{amssymb}
\usepackage{color}
\usepackage{balance} 
\usepackage{float}

\usepackage{makecell}
\usepackage{multirow}
\usepackage{diagbox}

\usepackage[bottom]{footmisc}

\usepackage{xparse}
\usepackage{tikz}
\usetikzlibrary{matrix,backgrounds}

\pgfdeclarelayer{myback}
\pgfsetlayers{myback,background,main}
\NewDocumentCommand{\ohighlight}{O{blue!40} m m}{%
\draw[mycolor=#1] (#2.north west)rectangle (#3.south east);
}
\NewDocumentCommand{\fhighlight}{O{blue!40} m m}{%
\draw[myfillcolor=#1] (#2.north west)rectangle (#3.south east);
}
\tikzset{mycolor/.style = {line width=1bp,color=#1}}%
\tikzset{myfillcolor/.style = {draw,fill=#1}}%

\newcommand{\bma}{\begin{bmatrix}}
\newcommand{\ebma}{\end{bmatrix}}

\usepackage{mathtools}

\newtheoremstyle{bfnote}%
{}{}%
{\itshape}{}%
{\bfseries}{.}%
{ }%
{\thmname{#1}\thmnumber{ #2}\thmnote{ (#3)}}
\theoremstyle{bfnote}

\usepackage{algorithm, algpseudocode}

\begin{document}

\title{Sensitivity-Based System Strength Assessment: Mapping Power Flow and Network Topology Perturbations to System Eigenvalues}

\author{Trager Joswig-Jones,~\IEEEmembership{Student Member,~IEEE,} Shuan Dong,~\IEEEmembership{Member,~IEEE,} Jin Tan,~\IEEEmembership{Member,~IEEE,}  Baosen Zhang,~\IEEEmembership{Member,~IEEE}
\thanks{T. Joswig-Jones and B. Zhang are with the Department of Electrical and Computer Engineering at the University of Washington, \{joswitra,zhangbao\}@uw.edu. J. Tan and S. Dong are with the National Laboratory of the Rockies \{shuan.dong,jin.tan\}@nlr.gov. T. Joswig-Jones and B. Zhang are supported in part by the Grainger Foundation and the Galloway Fellowships. This work was authored [in part] by the National Laboratory of the Rockies for the U.S. Department of Energy (DOE), operated under Contract No. DE-AC36-08GO28308. Funding provided by U.S. Department of Energy Office of Critical Minerals and Energy Innovation Integrated Energy Systems Office (\#52373). The views expressed in the article do not necessarily represent the views of the DOE or the U.S. Government. The U.S. Government retains and the publisher, by accepting the article for publication, acknowledges that the U.S. Government retains a nonexclusive, paid-up, irrevocable, worldwide license to publish or reproduce the published form of this work, or allow others to do so, for U.S. Government purposes.}
}



\maketitle

\begin{abstract}
As inverter-based resources (IBRs) contribute larger shares of generation in electrical power grids, quantifying system strength becomes increasingly important for identifying stability issues introduced by these devices. While admittance model based system strength metrics have been proposed to identify control interactions in systems with high levels of IBRs, these methods depend on repeated evaluations across many operating points to understand how the state of the system impacts system strength. To address this challenge, we consider the sensitivity of system stability to perturbations in the steady-state operating point, and propose system strength metrics based on sensitivities to power injections, voltages, and line admittances. Using these sensitivities we can identify changes in the system's state (e.g. a line tripping off or a generator increasing its power output) that trigger instability mechanisms. We show that these metrics provide critical insights into system stability and can be computed much faster than repeated eigenvalue calculations. We demonstrate our approach on 14-bus and 118-bus test systems to show how the metrics can be used to find remedial actions for small-signal stability issues.
\end{abstract}


\section{Introduction}
%
As conventional synchronous generation is replaced with inverter-based resources (IBRs) in electrical power systems, grid operators are faced with a range of new challenges when ensuring power systems can maintain stable voltage and frequency. 
In particular, the integration of grid-following (GFL) IBRs into power systems has made modern power systems increasingly prone to small-signal instabilities, such as subsynchronous oscillations~\cite{6344713, 9740416} which have been observed in power systems repeatedly over the past decade.
The ability of power systems to maintain stable voltage and frequency under disturbances can broadly be referred to as system strength~\cite{CIGRE2016}, and many metrics have been developed that quantify this system strength to identify weak areas of the grid and potential modes of instability.

While traditional indices like the short-circuit ratio (SCR) serve as standard screening tools, they rely on short-circuit analysis which cannot accurately quantify system strength for systems with high IBR penetration~\cite{CIGRE2016}. Consequently, they can fail to identify areas prone to instabilities caused by control interactions between IBR devices~\cite{Henderson2024}. To overcome these limitations, several metrics have been proposed, including admittance-based methods that directly consider the small-signal stability of the system~\cite{Joswig-Jones_Dong_Tan_2025,Ma_Xin_Wu_Liu_Chen_2024, Zhu_Green_Zhou_Li_Kong_Gu_2024}. In general, approaches that estimate small-signal stability regions~\cite{8939499, 9236196} or trace eigenvalue paths~\cite{bouterakos2025eigenvaluetrackinglargescalesystems} may not be practical or computationally feasible for screening large systems with IBRs; however, metrics based on the sensitivity of modes can inform operators on how to control the system to prevent small-signal instability.

To evaluate shifts in small-signal stability analytically, sensitivity-based approaches quantify how eigenvalues migrate in response to parameter variations. Previous studies have investigated methods for calculating the sensitivities to changes in device parameters~\cite{Zein1977, Smed1993, 852145}. However, calculating sensitivities with respect to power injections or bus voltages has not received much attention. The work in \cite{Ma_Dong_Zhang_2006} considered how to calculate eigenvalue sensitivities with respect to perturbations in power flow parameters, using white-box analytical device models.
To improve computational efficiency for large systems, modified numerical methods have been proposed to calculate eigenvalue sensitivities to power injections and voltages~\cite{Li2019_EfficientNumerical}. These finite-difference techniques require explicitly reconstructing the perturbed global state matrix for each parameter variation which again is not viable for systems containing proprietary, black-boxed IBRs~\cite{cifuentes2022blackbox,huang2026learning}. 
More recently, \cite{Zhu_Gu_Li_Green_2022} employed grey-box modeling techniques to compute the sensitivity of eigenvalues to changes in parameters, but again does not consider sensitivities to power flow parameters. The work in \cite{11168906} considers how changes in a device's control parameters impact its operating point, but only considered a single machine infinite bus setting.

Beyond operating point perturbations, the small-signal stability is also sensitive to topological changes such as line switching or unexpected line outages. Modifying the network's admittance structure forces a global redistribution of power flow, altering the steady-state operating points of all connected devices. In stressed or low-inertia grids, tripping a critical line can significantly diminish system strength or even drive weakly damped modes into instability~\cite{7038236, 9967438}. Therefore, identifying critical lines for contingency ranking analysis and understanding how the system's small-signal stability responds to structural network perturbations is essential for secure N-1 operation. 

\IEEEpubidadjcol  

In this paper, we define new system strength metrics that build on the work in~\cite{Zhu_Gu_Li_Green_2022, Zhu_Green_Zhou_Li_Kong_Gu_2024} to consider the sensitivity of the system eigenvalues to changes in power flow set points and line admittances. It extends the method in \cite{Joswig-Jones_Sensitivity_2026} by demonstrating how to calculate the sensitivities of system eigenvalues to perturbations in line admittances and considers how perturbations in the power flow impact the load models.  We demonstrate with 14-bus and 118-bus test systems, how such an approach can provide system operators with information to practically identify and mitigate small-signal stability challenges in real-time by redispatching generators in the system. In addition, we show that our method is much faster than conventional eigenvalue computations.  


Compared to \cite{Joswig-Jones_Sensitivity_2026}, the previous work presented how to evaluate eigenvalue sensitivities to perturbations in power injections, while this paper generalizes the approach to account for structural network perturbations. Specifically, the approach is extended to calculate eigenvalue sensitivities with respect to line admittances. This approach can also be used to consider the indirect impacts of power flow redistributions on equivalent load models. 

The remainder of this paper is organized as follows. 
Section~\ref{sec:model} introduces the system model. Section~\ref{sec:power-flow-sensitivity} discusses how to find the sensitivity of the power flow to perturbations in power flow parameters or line admittances and Section~\ref{sec:eigenvalue-perturbation} discusses how to use these sensitivities to calculate the sensitivities of the system's eigenvalues. Section~\ref{sec:pfs-metric} defines system strength metrics based on these power injection sensitivities and Section~\ref{sec:scalability} discusses their scalability. Section~\ref{sec:simulation} demonstrates the capabilities of these metrics with two test systems and Section~\ref{sec:conclusion} concludes the paper. 


\section{System Model}
\label{sec:model}
\begin{figure}
    \centering
    \includegraphics[width=\linewidth]{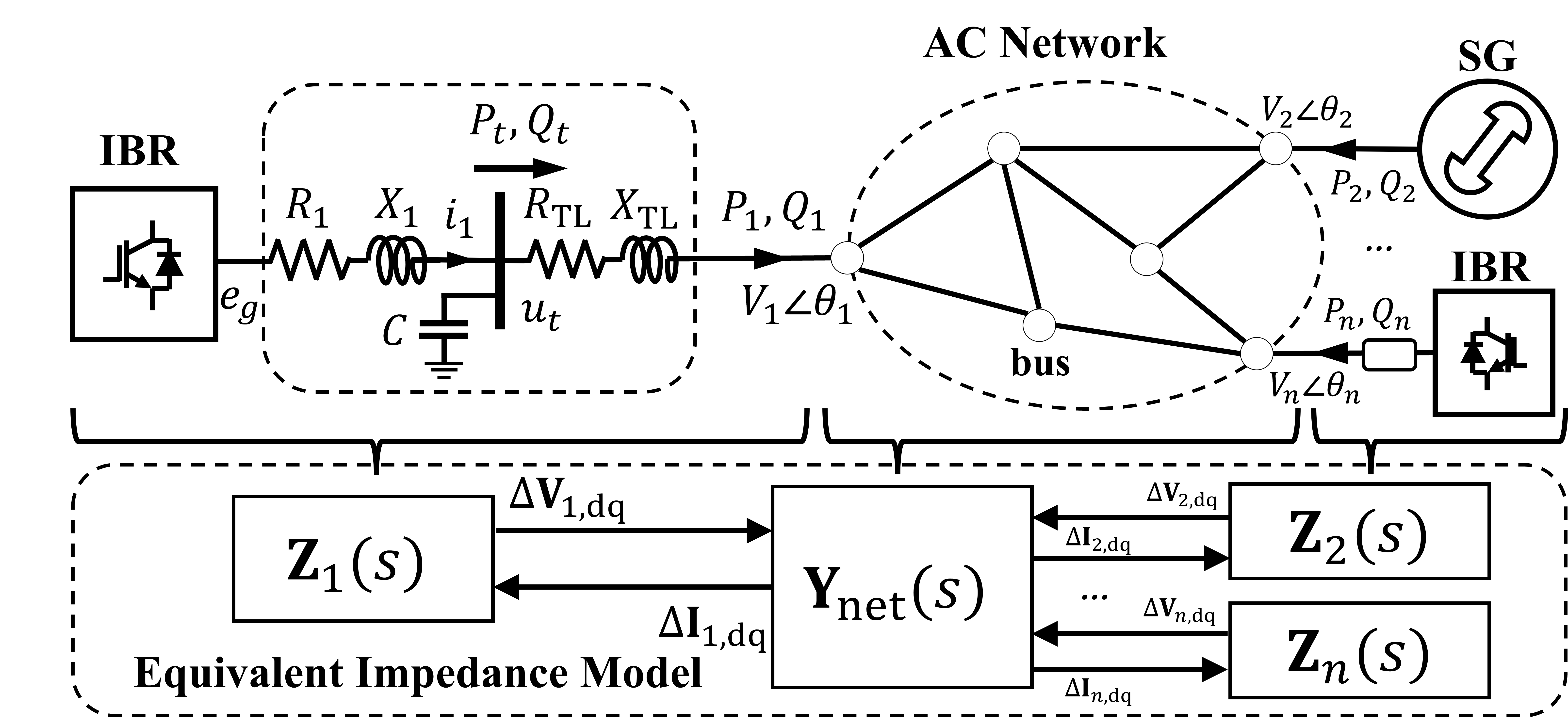}
    \caption{Illustration of a power system with its equivalent impedance model.}
    \label{fig:AcNetwork}
    \vspace{-5pt}
\end{figure}

We consider an AC power network as represented in Fig.~\ref{fig:AcNetwork}. 
The network comprises a set of $N$ buses, $\mathcal{N}$, and interconnected lines, $\mathcal{E}$, where $Y_{ik} = G_{ik} + j B_{ik}$ denotes the admittance between buses $i$ and $k$. We use $i\sim k$ to indicate the set of all buses $k$ that are directly connected to bus $i$.
Bus $i \in \mathcal{N}$ is characterized by power injections $S_i = P_i + j Q_i$ and voltage $V_i \angle \theta_i$; these are collected into vectors $\mathbf{V}, \mathbf{\Theta}, \mathbf{P}, \mathbf{Q}$.
$\theta_{ik}$ indicates the angle difference between bus $i$ and bus $k$ (i.e. $\theta_i - \theta_k)$.
A slack bus maintains system power balance by varying its power injections.

\subsection{Power Flow}

In steady-state, the power flow through the network satisfies the AC power flow equations
\begin{subequations}
\label{eq:pf}
\begin{align}
    P_i &=V_i \sum_{i \sim k}{V_k}(G_{ik} \cos{\theta_{ik}} + B_{ik} \sin{\theta_{ik})}, \\
    Q_i &=V_i \sum_{i \sim k}{V_k}(G_{ik} \sin{\theta_{ik}} - B_{ik} \cos{\theta_{ik})}, 
\end{align}
\end{subequations}
for each bus $i \in \mathcal{N}$. The first-order Taylor series expansion of (\ref{eq:pf}) gives 
\begingroup
\renewcommand*{\arraystretch}{1.2}
\begin{equation}
    \label{eq:first-order-pf}
    \begin{bmatrix}
    \Delta \bf{P} \\ \Delta \bf{Q}
    \end{bmatrix} = J(\mathbf{\mathcal{M}})
    \begin{bmatrix}
    \Delta \bf{V} \\ \Delta \bf{\Theta}
    \end{bmatrix}, \quad 
    J(\mathbf{\mathcal{M}})=\begin{bmatrix}
    \frac{\partial \bf{P}}{\partial \bf{V}} & \frac{\partial \bf{P}}{\partial \bf{\Theta}} \\ 
    \frac{\partial \bf{Q}}{\partial \bf{V}} & \frac{\partial \bf{Q}}{\partial \bf{\Theta}}
    \end{bmatrix},
\end{equation}
\endgroup
where $J(\mathcal{M})$ represents the unpartitioned Jacobian of the network power flow equations at the operating point $\mathcal{M}$. While standard load flow algorithms compute a portion of this Jacobian by omitting equations for constrained variables at PV and slack buses, $J$ contains the sensitivities across all $N$ buses. Here we denote the power flow parameters at all buses in the system as $\mathcal{M}$, where $\mathcal{M}_i = [P_i, Q_i, V_i, \Theta_i]$ are the parameters at bus $i$.

To solve power flow models, buses are specified as having two of their four power flow parameters being specified. Typically, bus types are \emph{PQ}, \emph{PV}, and \emph{V$\Theta$} (the slack bus) where the bus type indicates the parameters that are specified at that bus. 
We let $\overline{\mathcal{V}}$, $\overline{\Theta}$, $\overline{\mathcal{P}}$, and $\overline{\mathcal{Q}}$ denote the set of buses where the respective parameter is specified at that bus and 
we let $\widetilde{\mathcal{V}}$, $\widetilde{\Theta}$, $\widetilde{\mathcal{P}}$, and $\widetilde{\mathcal{Q}}$ denote the set of buses where the respective parameter value at that bus is dependent upon the specified values and (\ref{eq:pf}).

\subsection{Dynamics and Small-Signal Stability}
The dynamic behaviors of elements in the network are modeled via nonlinear state-space models in local $\mathrm{dq}$ reference frames. By linearizing these models around a specific power flow solution, local equivalent admittance/impedance matrices are derived, which are transfer functions relating voltage $\mathbf{V}_{i,\mathrm{dq}}$ and current $\mathbf{I}_{i,\mathrm{dq}}$ at the point of interconnection (POI). Aligning these models to a global dq frame allows for the aggregation of the whole-system admittance matrix, characterizing the global small-signal dynamics in the frequency domain~\cite{Gu_Li_Zhu_Green_2021}:
\begin{equation}
    \label{eq:whole-system-admittance}
    \hat{\mathbf{Y}}(s) = (\mathbf{I}+\mathbf{Y}_\mathrm{net}(s)\mathbf{Z}_\mathrm{dev}(s))^{-1} \mathbf{Y}_\mathrm{net}(s),
\end{equation}
where $\mathbf{Y}_\mathrm{net}(s)$ represents the network's nodal admittance matrix, and $\mathbf{Z}_\mathrm{dev}(s)=diag(\mathbf{Z}_1(s),\mathbf{Z}_2(s),\dots,\mathbf{Z}_K(s))$ constitutes a block-diagonal matrix of individual device impedances, and $\mathbf{I}$ is the identity matrix. 

The small-signal stability of the power system is dictated by the poles of $\hat{\mathbf{Y}}(s)$, which are mathematically equivalent to the system eigenvalues~\cite{Gu_Li_Zhu_Green_2021}. The system is stable if all eigenvalues remain in the left-hand plane. Furthermore, as established in~\cite{Zhu_Gu_Li_Green_2022}, a specific eigenvalue $\lambda$ exhibits a sensitivity to variations in the whole-system admittance matrix defined by $\frac{\partial\lambda}{\partial \hat{\mathbf{Y}}^{-1}(\lambda)} = -\mathrm{Res}_\lambda \hat{\mathbf{Y}}$. For sufficiently small topological or parametric perturbations, the resulting shift in the eigenvalue is computed as:
\vspace{-3pt}
\begin{equation}
    \label{eq:Dlambda}
    \Delta \lambda = \langle -\mathrm{Res}_{\lambda}^* \hat{\mathbf{Y}}, \Delta \hat{\mathbf{Y}}^{-1} \rangle,
\end{equation}
where $\langle \cdot,\cdot \rangle$ denotes the Frobenius inner product, $\mathrm{Res}_{\lambda} \hat{\mathbf{Y}}$ is the admittance matrix residue evaluated at pole $\lambda$, and $*$ indicates the conjugate transpose~\cite{stein2010complex}.

Crucially, because equivalent admittance modeling relies on the linearization of nonlinear components, the resulting transfer function models are intrinsically tied to the steady-state operating point. Any deviation in a device’s power injection or POI voltage shifts the point at which it is linearized, subsequently altering its equivalent impedance model and causing the system eigenvalues to shift.

\section{Method}
\label{sec:method}
\begin{figure*}
    \centering
    \includegraphics[width=0.85\textwidth]{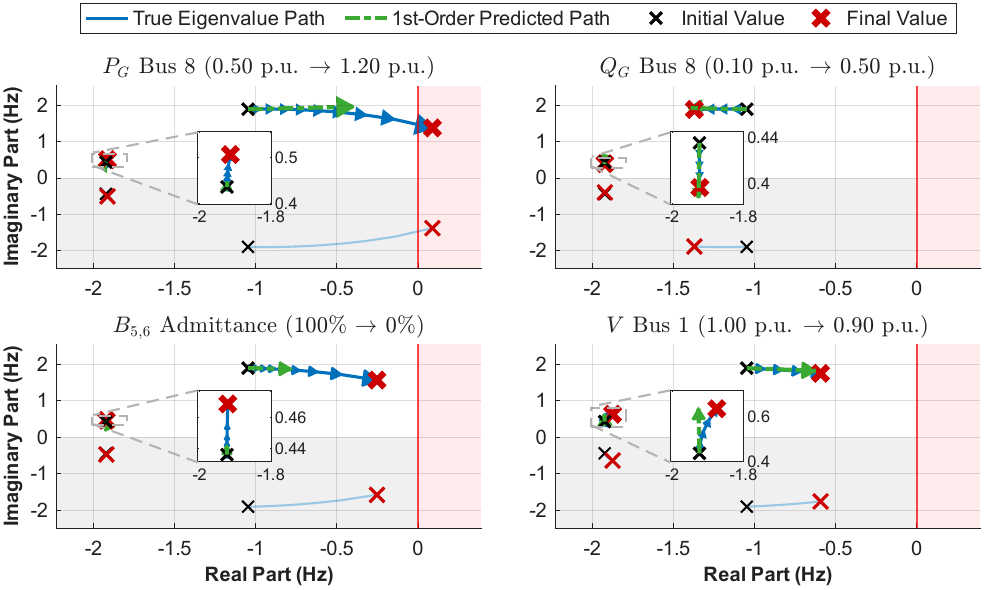}
    \caption{14-bus case select eigenvalue traces with first-order sensitivity perturbation prediction for changes in power flow and network admittance parameters.
    }
    \label{fig:eig-traces}
    \vspace{-5pt}
\end{figure*}

The proposed methodology uses analytical sensitivities to evaluate how system eigenvalues respond to power flow parameter changes. As demonstrated in Fig.~\ref{fig:eig-traces}, while first-order approximations do not directly capture highly nonlinear trajectories under large parameter variations, they successfully indicate the initial direction and magnitude of modal perturbations. To formulate these metrics, this section first evaluates how parameter changes shift the operating point, and subsequently maps those shifts to the system's global eigenvalues. The resulting sensitivities are used to define system strength metrics. 

%
%


\subsection{Power Flow Sensitivity}
\label{sec:power-flow-sensitivity}
%
\subsubsection{Sensitivity to \texorpdfstring{$P, Q, V$}{P, Q, V}} 
\label{sec:power-flow-param-sensitivity}
When the power injection or voltage magnitude at a bus changes, the system adjusts to a new power flow solution satisfying (\ref{eq:pf}). 
Accurately assessing the impact of such a perturbation on small-signal stability requires accounting for its effect on the overall power flow as this determines the operating point the devices in the system are linearized about~\cite{Ma_Dong_Zhang_2006}. Although this nonlinear problem is generally challenging, for sufficiently small perturbations it can be approximated using a first-order expansion. In this section we discuss how to calculate the power flow perturbation $\Delta \mathcal{M}$ for a perturbation to a chosen specified power flow parameter $\Delta x$ (e.g. for a perturbation in the active power at bus 2 set $x = P_2$). For notational simplicity hereafter, the explicit operating point is omitted, such that $J \equiv J(\mathcal{M})$ represents the Jacobian at the base power flow. 

First, we partition and rearrange the Jacobian relating to whether or not a power flow parameter is specified or dependent and define sub-matrices as shown in (\ref{eq:Jacobian-Partition}).

\begin{equation}
\label{eq:Jacobian-Partition}
\hbox{

$J_{\bullet,\bullet}=$
\begin{tikzpicture}[baseline=(current  bounding  box.center)]
\matrix (m)[matrix of math nodes,left delimiter={[},right delimiter={]}]
{
        \frac{\partial \overline{\mathbf{P}}}{\partial \overline{\mathbf{V}}} & \frac{\partial \overline{\mathbf{P}}}{\partial \overline{\mathbf{\Theta}}} & 
        \frac{\partial \overline{\mathbf{P}}}{\partial \widetilde{\mathbf{V}}} & \frac{\partial \overline{\mathbf{P}}}{\partial \widetilde{\mathbf{\Theta}}}\\ 
        \frac{\partial \overline{\mathbf{Q}}}{\partial \overline{\mathbf{V}}} & \frac{\partial \overline{\mathbf{Q}}}{\partial \overline{\mathbf{\Theta}}} & 
        \frac{\partial \overline{\mathbf{Q}}}{\partial \widetilde{\mathbf{V}}} & \frac{\partial \overline{\mathbf{Q}}}{\partial \widetilde{\mathbf{\Theta}}}\\ 
        \frac{\partial \widetilde{\mathbf{P}}}{\partial \overline{\mathbf{V}}} & \frac{\partial \widetilde{\mathbf{P}}}{\partial \overline{\mathbf{\Theta}}} & 
        \frac{\partial \widetilde{\mathbf{P}}}{\partial \widetilde{\mathbf{V}}} & \frac{\partial \widetilde{\mathbf{P}}}{\partial \widetilde{\mathbf{\Theta}}}\\ 
        \frac{\partial \widetilde{\mathbf{Q}}}{\partial \overline{\mathbf{V}}} & \frac{\partial \widetilde{\mathbf{Q}}}{\partial \overline{\mathbf{\Theta}}} & 
        \frac{\partial \widetilde{\mathbf{Q}}}{\partial \widetilde{\mathbf{V}}} & \frac{\partial \widetilde{\mathbf{Q}}}{\partial \widetilde{\mathbf{\Theta}}}\\ 
};

\begin{pgfonlayer}{myback}
\fhighlight[blue!30]{m-1-1}{m-2-2}
\fhighlight[green!30]{m-1-3}{m-2-4}
\fhighlight[red!20]{m-3-1}{m-4-2}
\fhighlight[violet!20]{m-3-3}{m-4-4}
\end{pgfonlayer}
\end{tikzpicture}
$=$  
\begin{tikzpicture}[baseline=(current  bounding  box.center)]
\matrix (m)[matrix of math nodes,left delimiter={[},right delimiter={]}]
{
        J_{-,-} & J_{-,\sim} \\ 
        J_{\sim,-} & J_{\sim,\sim} \\ 
};

\begin{pgfonlayer}{myback}
\fhighlight[blue!30]{m-1-1}{m-1-1}
\fhighlight[green!30]{m-1-2}{m-1-2}
\fhighlight[red!20]{m-2-1}{m-2-1}
\fhighlight[violet!20]{m-2-2}{m-2-2}
\end{pgfonlayer}
\end{tikzpicture}

}
\end{equation}
Here $J_{-,\sim}$ represents the mapping of dependent voltage parameters, $(\widetilde{\mathcal{V}},\widetilde{\Theta})$, to specified power parameters, $(\overline{\mathcal{P}},\overline{Q})$. We note that $J_{-,\sim}$ is the matrix used in standard Newton-Raphson AC power flow algorithms, and it is invertible as long as the system is not at a point of voltage collapse and the network is topologically connected.
We further define $J_{-,\bullet}= [J_{-,-}, J_{-, \sim}]$, which selects the rows of $J$ corresponding to specified power injections and the columns corresponding to all voltage values.

For a perturbation in the active power at buses in $\overline{\mathcal{P}}$ or reactive power at buses in $\overline{\mathcal{Q}}$ ($x = P_i, i \in \overline{\mathcal{P}}$ or $x = Q_i, i \in \overline{\mathcal{Q}}$) we can calculate the voltage magnitude and angle sensitivities at buses in $\widetilde{\mathcal{V}}$ and $ \widetilde{\Theta}$, respectively, from 
\begin{equation}
    \label{eq:pqbarfromvth}
    \left(
        \frac{\partial {\mathbf{P}}}{\partial x },
        \frac{\partial {\mathbf{Q}}}{\partial x }
    \right) =
    J \cdot
    \left(
        \frac{\partial {\mathbf{V}}}{\partial x } \\
        \frac{\partial {\mathbf{\Theta}}}{\partial x }
    \right)
\end{equation}
%
%
by taking $\frac{\partial \overline{\mathbf{V}}}{\partial x}, \frac{\partial \overline{\mathbf{\Theta}}}{\partial x} = \mathbf{0}$ to get
\begingroup
\renewcommand*{\arraystretch}{1.5}
\begin{equation}
    \label{eq:vth-from-pqbar}
    \begin{bmatrix} 
        \frac{\partial \widetilde{\mathbf{V}}}{\partial x } \\
        \frac{\partial \widetilde{\mathbf{\Theta}}}{\partial x }
    \end{bmatrix} = 
    J_{-,\sim}^{-1}
    \begin{bmatrix} 
        \frac{\partial \overline{\mathbf{P}}}{\partial x } \\
        \frac{\partial \overline{\mathbf{Q}}}{\partial x }
    \end{bmatrix}.
\end{equation}
\endgroup
We can then calculate the sensitivities of $\widetilde{\mathbf{P}}, \widetilde{\mathbf{Q}}$ as 
\begingroup
\renewcommand*{\arraystretch}{1.5}
\begin{equation}
    \label{eq:vth2pqtilde}
    \begin{bmatrix} 
        \frac{\partial \widetilde{\mathbf{P}}}{\partial x } \\
        \frac{\partial \widetilde{\mathbf{Q}}}{\partial x }
    \end{bmatrix} = 
    J_{\sim,\bullet}
    \begin{bmatrix} 
        \frac{\partial \mathbf{V}}{\partial x } \\
        \frac{\partial \mathbf{\Theta}}{\partial x }
    \end{bmatrix}.
\end{equation}
\endgroup
We combine these sensitivities to get the local sensitivities $\frac{\partial \mathcal{M}_i}{\partial x} = [\frac{\partial{P}_i}{\partial x}, \frac{\partial {Q}_i}{\partial x}, \frac{\partial {V}_i}{\partial x}, \frac{\partial {\Theta}_i}{\partial x}]$ at each bus in the system.

Similarly, for a perturbation in $\overline{\mathbf{V}}$ or $\overline{\mathbf{\Theta}}$ values, we can calculate the sensitivities such that $\frac{\partial \overline{\mathbf{P}}}{\partial x}, \frac{\partial \overline{\mathbf{Q}}}{\partial x} = \mathbf{0}$ from~\eqref{eq:pqbarfromvth} to find $\frac{\partial \widetilde{\mathbf{V}}}{\partial x}, \frac{\partial \widetilde{\mathbf{\Theta}}}{\partial x}$ as
\begingroup
\renewcommand*{\arraystretch}{1.5}
\begin{equation}
    \label{eq:vthtilde-from-vthbar}
    \begin{bmatrix} 
        \frac{\partial \widetilde{\mathbf{V}}}{\partial x } \\
        \frac{\partial \widetilde{\mathbf{\Theta}}}{\partial x }
    \end{bmatrix}
    = -J_{-,\sim}^{-1} \cdot J_{-,-}
    \begin{bmatrix} 
    \frac{\partial \overline{\mathbf{V}}}{\partial x } \\
    \frac{\partial \overline{\mathbf{\Theta}}}{\partial x }
\end{bmatrix}.
\end{equation}
\endgroup
Then we can find $\frac{\partial \widetilde{\mathbf{P}}}{\partial x}, \frac{\partial \widetilde{\mathbf{Q}}}{\partial x}$ from~(\ref{eq:vth2pqtilde}).

Note that these first-order power flow sensitivity calculations are linear, such that we can combine the sensitivities from multiple power flow parameter perturbations by summing the sensitivity values for each individual power flow parameter perturbation.



\subsubsection{Sensitivity to Line Admittances} 
\label{sec:power-flow-line-sensitivity}

A change in a network line admittance value will also result in a change in the power flow values. Assuming that the line admittance is perturbed such that $\tau$, the $R/L$ line impedance ratio, is held constant then from $G_{ik} + j B_{ik} = \frac{R_{ik} - jX_{ik}}{R_{ik}^2 + X_{ik}^2}$ we have $B_{ik} = \frac{1}{X_{ik} (1 + (\tau / \omega_\mathrm{nom})^2)}, G_{ik} = \tau / \omega_\mathrm{nom} \cdot B_{ik}$. The first-order sensitivity of the dependent power injection values to a change in this admittance can be calculated from (\ref{eq:pf}) for bus i by taking the partial derivatives
\begingroup
\renewcommand*{\arraystretch}{1.55}
\begin{equation}
\label{eq:networkpfperturb}
\begin{bmatrix}
    \frac{\partial P_i}{\partial B_{ik}} \\
    \frac{\partial Q_i}{\partial B_{ik}}
\end{bmatrix} =
\begin{bmatrix}
    V_i^2 \left(\frac{\tau}{\omega_\mathrm{nom}}\right) - V_i V_k \left( \frac{\tau}{\omega_\mathrm{nom}} \cos{\theta_{ik}} - \sin{\theta_{ik}} \right) \\
    V_i^2 - V_i V_k \left( \frac{\tau}{\omega_\mathrm{nom}} \sin{\theta_{ik}} + \cos{\theta_{ik}} \right)
\end{bmatrix}
\end{equation}
\endgroup
Swapping the indices in \eqref{eq:networkpfperturb} we can calculate the power flow sensitivities at the other bus $k$ that the line connects to. Taking these sensitivities, we define 
$\mathbf{b} = ({\partial \mathbf{P}}/{\partial B_{ik}}, \; {\partial \mathbf{Q}}/{\partial B_{ik}})$
The vector $\mathbf{b}$ is accordingly partitioned into $\overline{\mathbf{b}}$ (corresponding to the specified injections) and $\widetilde{\mathbf{b}}$ (corresponding to the dependent injections). Because the specified power parameters are held constant during the line perturbation, their variations are zero.

Applying these boundary conditions, the first-order Taylor expansion of the power flow equations reduces to:
\begingroup
\renewcommand*{\arraystretch}{1.5}
\begin{equation}
    \mathbf{0} = J_{-, \sim} 
    \begin{bmatrix}
        \frac{\partial \widetilde{\mathbf{\Theta}}}{\partial B_{ik}} \\
        \frac{\partial \widetilde{\mathbf{V}}}{\partial B_{ik}}
    \end{bmatrix}
    + \overline{\mathbf{b}}
\end{equation}
\endgroup
where $J_{-, \sim}$ is the sub-matrix of the partitioned power flow Jacobian containing the partial derivatives of the specified injections with respect to the unknown state variables. 

The sensitivities of the unknown voltage magnitudes and angles across the network can therefore be solved directly as
\begingroup
\renewcommand*{\arraystretch}{1.5}
\begin{equation}
    \label{eq:state_sens}
    \begin{bmatrix}
        \frac{\partial \widetilde{\mathbf{\Theta}}}{\partial B_{ik}} \\
        \frac{\partial \widetilde{\mathbf{V}}}{\partial B_{ik}}
    \end{bmatrix} =
    - J_{-, \sim}^{-1} \overline{\mathbf{b}}.
\end{equation}
\endgroup

Finally, the sensitivities of the dependent power injections can be evaluated using the Jacobian sub-matrix $J_{\sim, \sim}$ corresponding to those dependent variables:
\begingroup
\renewcommand*{\arraystretch}{1.5}
\begin{equation}
    \label{eq:dep_inj_sens}
    \begin{bmatrix}
        \frac{\partial \widetilde{\mathbf{P}}}{\partial B_{ik}} \\
        \frac{\partial \widetilde{\mathbf{Q}}}{\partial B_{ik}}
    \end{bmatrix} =
    J_{\sim, \sim}
    \begin{bmatrix}
        \frac{\partial \widetilde{\mathbf{\Theta}}}{\partial B_{ik}} \\
        \frac{\partial \widetilde{\mathbf{V}}}{\partial B_{ik}}
    \end{bmatrix}
    + \widetilde{\mathbf{b}}.
\end{equation}
\endgroup

%

\subsection{Eigenvalue Sensitivity}
\label{sec:eigenvalue-perturbation}
%
\subsubsection{Sensitivities to Power Flow}
Using these power flow sensitivities we can determine the system eigenvalue sensitivities using intermediate device impedance sensitivity to changes in the local power flow parameters.
This requires the sensitivities $\frac{\partial Z_{k}}{\partial\mu}$ of the device impedances with respect to local power flow parameters. Equivalent device impedances and these sensitivities can be obtained from analytical models or from frequency-scanning measurements, as described in~\cite{Zhu_Gu_Li_Green_2022}. Because most black-box inverter models allow specification of power references, this approach is also compatible with black-box modeling when using frequency scanning to obtain device admittance models. 

Using the chain rule, we can determine the sensitivity of a device's equivalent impedance to a power flow parameter $x$ as 
\begin{equation}
\label{eq:deviceZ_dx}
\frac{\partial \mathbf{Z}_k}{\partial x} = \sum_{\mu \in \mathcal{M}_k} \frac{\partial \mathbf{Z}_k}{\partial \mu} \frac{\partial \mu}{\partial x}.
\end{equation}
We collect these as $\frac{\partial \mathbf{Z}_\mathrm{dev}}{\partial x} = \mathrm{diag}(\frac{\partial \mathbf{Z}_1}{\partial x}, \frac{\partial \mathbf{Z}_2}{\partial x}, \dots,\frac{\partial \mathbf{Z}_K}{\partial x})$. Then using~(\ref{eq:whole-system-admittance}) and~(\ref{eq:Dlambda}), noting that $\frac{\partial \hat{\mathbf{Y}}^{-1}}{\partial x} = \frac{\partial{\mathbf{Z}_\mathrm{dev}}}{\partial x}$ (See Appendix~\ref{app:proof} for details) we calculate the sensitivity of the system eigenvalue $\lambda$ as
\begin{equation}
    \label{eq:Dlambda-DZpf}
    \frac{\partial \lambda}{\partial x} = \left\langle -\mathrm{Res}_{\lambda}^* \hat{\mathbf{Y}}, \frac{\partial \mathbf{Z}_\mathrm{dev}}{\partial x} \right\rangle.
\end{equation}
Since ${\mathbf{Z}_\mathrm{dev}}$ is a block-diagonal matrix, this formulation can be localized to
$
    \label{eq:Dlambda-DZpf-k}
    \frac{\partial \lambda}{\partial x} = \sum_{k \in \mathcal{N}}\langle -\mathrm{Res}_{\lambda}^* \hat{\mathbf{Y}}_{kk}, \frac{\partial \mathbf{Z}_k}{\partial x} \rangle,
$
where $\mathbf{\hat{Y}}_{kk}$ are $2\times2$ diagonal blocks of $\mathbf{\hat{Y}}$ corresponding to bus $k$, which can be calculated from the local device impedance and the grid impedance seen from that device's POI~\cite{Zhu_Gu_Li_Green_2022}.

\subsubsection{Sensitivities to Line Admittance}

Perturbing the admittance of a line, not only impacts the device equivalent admittances through perturbations to the power flow parameters, but it also directly impacts the eigenvalues through the perturbation to the network admittance transfer function matrix $\mathbf{Y}_\mathrm{net}$. The sensitivity of the network admittance $\mathbf{Y}_\mathrm{net}$ to a specific line susceptance $B_{ij}$ is
\begin{equation}
    \label{eq:netperturbfromline}
     \frac{\partial \mathbf{Y}_\mathrm{net}(s)}{\partial B_{ij}} = \mathcal{L}_{ij} \otimes 
    \begin{bmatrix}
        \beta(s) & \alpha(s) \\
        -\alpha(s) & \beta(s)
    \end{bmatrix}
\end{equation}
where $\mathcal{L}_{ij}$ is the Laplacian matrix with only the entries for the line being perturbed, $B_{ij}$ is the admittance of the line, $\otimes$ represent the Kronecker product, and $\alpha(s) = \frac{\omega_0}{(s + \tau)^2 / \omega_0 + \omega_0}, \beta(s) = \frac{s + \tau}{(s + \tau)^2 / \omega_0 + \omega_0}$. 
Then using~(\ref{eq:whole-system-admittance}) and~(\ref{eq:Dlambda}), we calculate the eigenvalue sensitivity as
\begin{equation}
    \label{eq:Dlambda-DBij}
    \frac{\partial\lambda}{\partial B_{ij}} = \left\langle -\mathrm{Res}_{\lambda}^* \hat{\mathbf{Y}}, \mathbf{Y}_\mathrm{net}^{-1} \frac{\partial \mathbf{Y}_\mathrm{net}(s)}{\partial B_{ij}} \mathbf{Y}_\mathrm{net}^{-1} + \frac{\partial \mathbf{Z}_\mathrm{dev}}{\partial B_{ij}} \right\rangle.
\end{equation}
where $\frac{\partial \mathbf{Y}_\mathrm{net}(s)}{\partial B_{ij}}$ comes from \eqref{eq:netperturbfromline} and $\frac{\partial \mathbf{Z}_\mathrm{dev}}{\partial B_{ij}}$ is obtained by calculating \eqref{eq:deviceZ_dx} for each device using the power flow sensitivities described in Section~\ref{sec:power-flow-line-sensitivity}.
The derivation of this can be found in Appendix~\ref{app:netproof}. 

\subsubsection{Sensitivities of Load Models}

In small-signal stability analysis, static constant power loads are modeled as equivalent parallel shunts. Linearized around a steady-state operating point, the equivalent shunt admittance $Y_{Lk} = G_{Lk} + jB_{Lk}$ at bus $k$ is calculated from the specified power demand ($P_{Lk}, Q_{Lk}$) and steady-state voltage magnitude $V_k$ as:
$
    G_{Lk} = {P_{Lk}}/{V_k^2}, \: B_{Lk} = -{Q_{Lk}}/{V_k^2}.
$
Loads can also be modeled as ZIP loads, where portions of the demand are classified as constant impedance, constant current, and constant power. The selection of these ZIP proportions impacts the damping of the system; constant impedance components increase system damping, whereas constant power components introduce negative incremental resistance that reduces damping.

These equivalent load admittances are embedded within the network matrix $\mathbf{Y}_\mathrm{net}(s)$. Therefore, a change in a local power flow parameter $x$ induces a sensitivity in the network matrix relating to the load:
\begin{equation}
    \label{eq:network-line-sensitivity}
    \frac{\partial \mathbf{Y}_\mathrm{net}(s)}{\partial x} = E_{kk} \otimes \frac{\partial L_{kk}(s)}{\partial x}
\end{equation}
where $E_{kk}$ is an $N \times N$ standard basis matrix with a $1$ at the $k$-th diagonal, and $L_{kk}(s)$ is the $2 \times 2$ admittance transfer function matrix of the load. $\frac{\partial L_{kk}(s)}{\partial x}$ is found by differentiating $G_{Lk}$ and $B_{Lk}$ with respect to $x$.

The eigenvalue sensitivity to this load perturbation is then evaluated using the network sensitivity formulation in (\ref{eq:Dlambda-DBij}). For a structural network perturbation (like a line outage) that simultaneously alters load voltages, the total $\frac{\partial \mathbf{Y}_\mathrm{net}}{\partial B_{ij}}$ is the sum of the direct line contribution from (\ref{eq:netperturbfromline}) and this indirect load contribution.

Note that loads can also be represented with dynamic models (e.g., induction motors or inverters) rather than static steady-state equivalents. In such cases, the dynamic load is modeled as a standard power electronic or electromechanical apparatus, and its power-flow-driven impedance sensitivities are incorporated into the device matrix $\mathbf{Z}_\mathrm{dev}(s)$ and evaluated using the method described in Section (\ref{eq:Dlambda-DZpf-k}).


\subsection{Small-Signal Stability Metrics} 
\label{sec:pfs-metric}
%
In this section we define system strength metrics based on the first-order approximation of eigenvalue shifts to changes in system parameters. Using the eigenvalue sensitivity $\frac{\partial \lambda}{\partial \mu}$ we can calculate a scaled eigenvalue perturbation $\Delta_\mu \lambda$ as
$$
\Delta_\mu \lambda = \frac{\partial \lambda}{\partial \mu} \Delta \mu,
$$
where $\Delta \mu$ can reflect the sensitivity to an absolute change in the parameter ($\Delta \mu = 1~\mathrm{pu}$) or a proportion of the base parameter value ($\Delta \mu = \mu_0$). We define the sensitivity index for an eigenvalue $\lambda$ and parameter $\mu$ at bus $i$ as
\begin{equation}
\mu\text{SI}_{i,\lambda} = \frac{\mathrm{re}(\Delta_{\mu} \lambda)}{|\mathrm{re}(\lambda)|}.
\end{equation}
This normalizes the sensitivity of the real part of an eigenvalue by the magnitude of the real part of that eigenvalue. The smaller the real part of an eigenvalue and the higher its sensitivity, the larger the sensitivity index will be. From this we can consider the maximum normalized eigenvalue shift towards the right hand plane for an increase (to) or decrease (from) in the power value and define the to sensitivity index ($\mu$TSI) and from sensitivity index ($\mu$FSI) as
\begin{align}
\mu\text{TSI}_i = \max_{\lambda} (\mu\text{SI}_{i,\lambda}), \quad \mu\text{FSI}_i = \max_{\lambda} (-\mu\text{SI}_{i,\lambda}). 
\end{align}
The inverse of the $\mu$TSI and $\mu$FSI sensitivity indices give a first-order approximate of how large a perturbation to the parameter can be before the eigenvalue will cross into the RHP and the system will become unstable. With this intuition, it is clear that larger sensitivities indicates a weaker system which we show in the case study results in the following section. 


\subsection{Calculation Scalability}
\label{sec:scalability}

This sensitivity-based approach is significantly faster than recalculating full system eigenvalues at perturbed operating points, as it leverages system residues to compute sensitivities through simple linear algebraic operations. 

From an implementation perspective, acquiring equivalent admittance models and their sensitivities from black-box models through vector fitting~\cite{772353} is the most intensive step; however, these device-level admittances are independent and can be calculated in parallel or pre-calculated offline for a range of operating points. 

When evaluating large-scale systems with high state dimensionality, specific computational strategies are required. Large time-scale separation in power systems yields highly stiff state matrices, while the dense clustering of network modes causes their corresponding eigenvectors to become nearly parallel. Together, these factors generate an ill-conditioned global eigenbasis that induces severe numerical instability and precision loss in standard dense eigensolvers~\cite{Golub2013, Kundur1994}. To overcome this, selective eigenvalue algorithms~\cite{demmel1997applied_ch7, Rommes2008} can be utilized to project the system into targeted frequency bands, allowing for the accurate extraction of critical eigen-pairs. Furthermore, sparse state-space representations are employed to avoid the numerical precision challenges inherent to high-order transfer function matrices.

\begin{figure}
    \centering
    \includegraphics[width=0.9\linewidth]{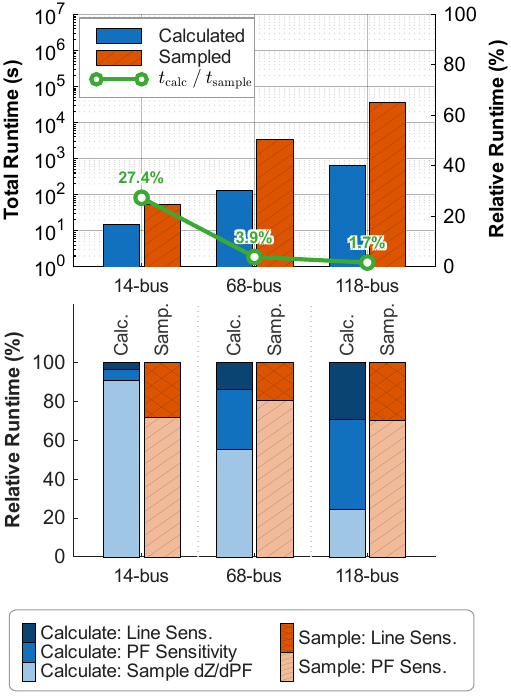}
    \caption{Runtimes comparing calculating the eigenvalue sensitivities to resampling the eigenvalues for a perturbation in a parameter for a range of systems. $P_G, Q_G, V$ sensitivities are tested for all specified power flow values at buses, and the $B_{ij}$ sensitivities are tested for all non-bridging lines.
    }
    \label{fig:sens-runtimes}
    \vspace{-3pt}
\end{figure}

The computational scalability of the proposed framework is demonstrated with the systems described in Table~\ref{tab:test-system-parameters}. Fig.~\ref{fig:sens-runtimes} compares the total runtimes for calculating eigenvalue sensitivities using our analytical approach against iterative numerical sampling. The total runtimes encompass evaluations for all specified power flow parameters and non-bridging line admittances across a designated set of critical eigenvalues. As the system size grows, the analytical approach yields an increasingly significant reduction in overall computation time.

\setlength{\tabcolsep}{3pt} 
\renewcommand{\arraystretch}{1.0} 
\begin{table}
    \centering
    \caption{Test System Parameters}
    \begin{tabular}{c|l l l l c c}
       \textbf{System}  &  \textbf{Buses} &  \textbf{Lines} &  \textbf{Devices} &  \textbf{States} &  \textbf{\thead{Considered \\ Eigenvalues}} & \textbf{\thead{Considered \\ Lines}} \\ \hline
       14-bus &  14 & 6  & 20 & 249 & 16 & 19 \\ \hline
       68-bus &  68 & 83 & 29 & 1041 & 35 & 65 \\ \hline
       118-bus & 118 & 179 & 26 & 2130 & 66 & 170 \\ \hline
    \end{tabular}
    \label{tab:test-system-parameters}
\end{table}

\section{Case Study} 
\label{sec:simulation}
%
To validate the proposed metrics in identifying weak buses and informing remedial actions, case studies are performed using a modified version of the Simplus Grid Tool~\cite{simplusgt} (source code will be made available at \url{https://github.com/TragerJoswig-Jones/Simplus-Grid-Tool-System-Sensitivity}).

\subsection{14-Bus Test System Results}
\begin{figure}
\vspace{-10pt}
    \centering
    \includegraphics[width=0.8\linewidth,trim={15 15 15 5},clip]{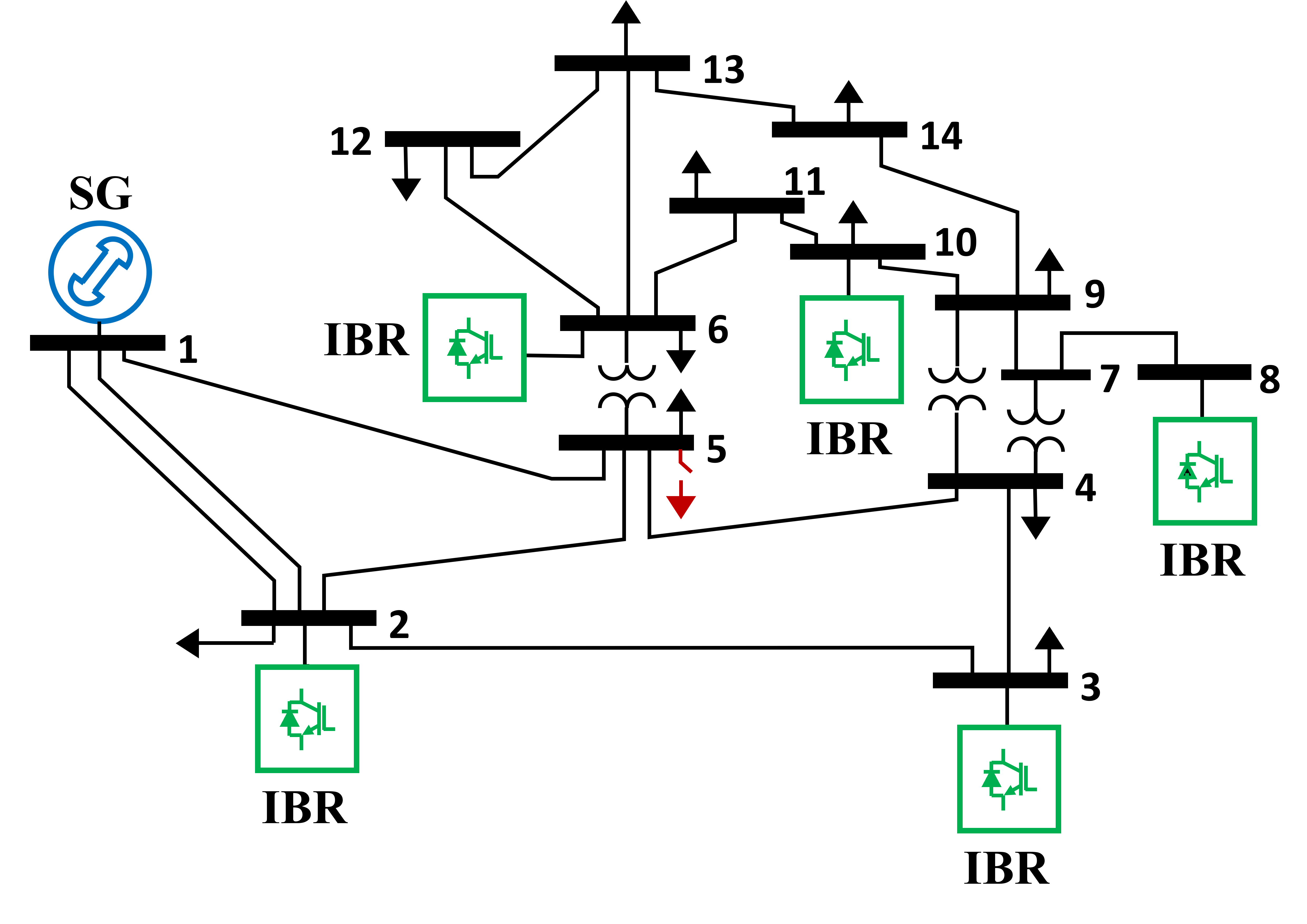}
    \vspace{-5pt}
    \caption{Diagram of the modified IEEE 14-bus test system.}
    \label{fig:IEEE14bus}
    \vspace{-3pt}
\end{figure}

First, we evaluate a modified IEEE 14-bus system (Fig.~\ref{fig:IEEE14bus}) where synchronous condensers are replaced by identical GFL IBRs, with one additional IBR at bus 10. All non-slack buses are modeled as \emph{PQ} buses. Table~\ref{tab:eig-perturbations} verifies the accuracy of the proposed analytical derivations against recalculating perturbed eigenvalues, demonstrating close alignment across active power, reactive power, voltage, and admittance perturbations.
\setlength{\tabcolsep}{3pt} 
\renewcommand{\arraystretch}{1.0} 
\begin{table}
    \centering
    \caption{IEEE 14-bus case study case descriptions}
    \begin{tabular}{c|l}
       \textbf{Case}  &  \textbf{Changes from Base Case} \\ \hline
       Case A~~ & $P_G: 0.5 \to 1.2~\text{pu}$ at bus \textbf{8} and bus \textbf{10} \\ \hline
       Case B~~ & $P_G: 0.5 \to 1.2~\text{pu}$ at bus \textbf{6} and bus \textbf{10} \\ \hline
       Case A-2 & \makecell[l]{$P_G: 0.5 \to 1.2~\text{pu}$ at bus \textbf{8} and bus \textbf{10}, \\ $Q_G\hspace{-0.2em}: 0.05 \to 0.5~\text{pu}$ at bus \textbf{8}.}  \\ \hline
       Case A-3 & \makecell[l]{$P_G: 0.5 \to 1.2~\text{pu}$ at bus \textbf{8} and bus \textbf{10}, \\ Bus \textbf{8} GFL to GFM control.} \\
    \end{tabular}
    \label{tab:case-descriptions}
    \vspace{-10pt}
\end{table}
\begin{figure}
    \centering
    \includegraphics[width=0.7\linewidth]{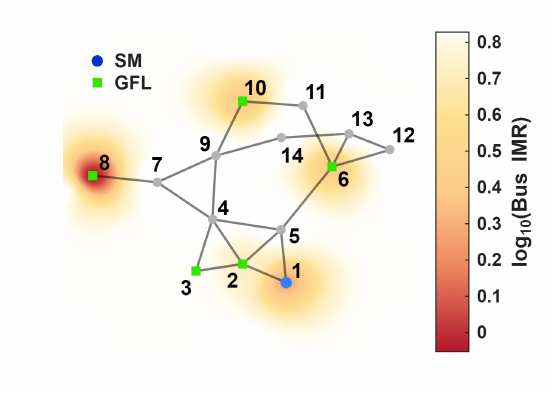}
    \caption{14-bus system base case IMR values.}
    \label{fig:14busIMR}
    \vspace{-12pt}
\end{figure}
\begin{figure*}[!ht]
    \centering
    \begin{subfigure}[t]{0.20\textwidth}
        \centering
        \includegraphics[trim={15 15 57 15},clip,height=\textwidth]{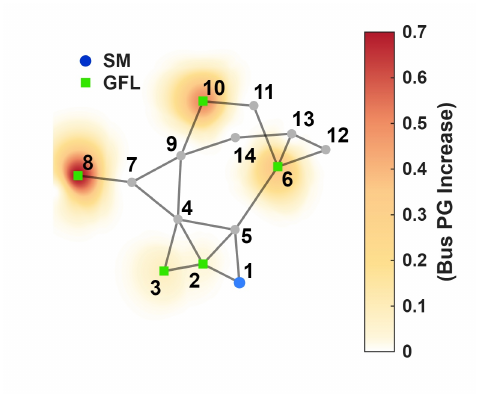}
        \caption{PTSI}
    \end{subfigure}%
    ~ 
    \begin{subfigure}[t]{0.20\textwidth}
        \centering
        \includegraphics[trim={15 15 57 15},clip, height=\textwidth]{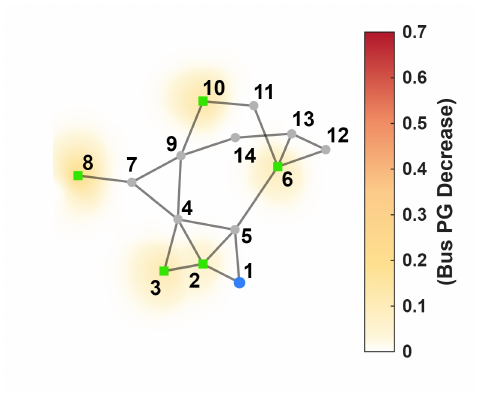}
        \caption{PFSI}
        \label{fig:PbusScale}
    \end{subfigure}
    ~
    \begin{subfigure}[]{0.05\textwidth}
        \centering
        \hspace{-10pt}
        \vspace{10pt}
        \includegraphics[trim={170 15 24 11},clip,height=4\textwidth]{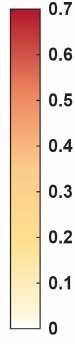}
        \vspace{85pt}
    \end{subfigure}
    \vspace{-40pt}
    ~
    \begin{subfigure}[t]{0.20\textwidth}
        \centering
        \includegraphics[trim={15 15 57 15},clip,height=\textwidth]{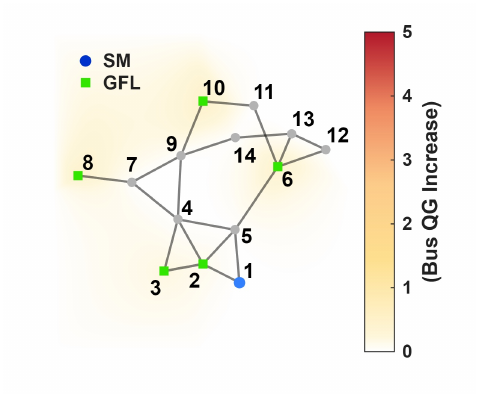}
        \caption{QTSI}
        \label{fig:14busQtosens}
    \end{subfigure}
    ~
    \begin{subfigure}[t]{0.20\textwidth}
        \centering
        \includegraphics[trim={15 15 57 15},clip,height=\textwidth]{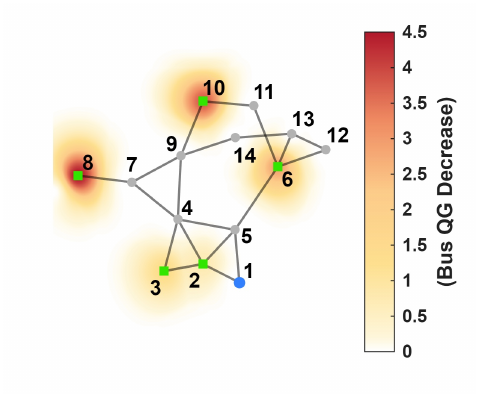}
        \caption{QFSI}
        \label{fig:14busQfrsens}
    \end{subfigure}
    ~
    \begin{subfigure}[]{0.05\textwidth}
        \centering
        \hspace{-10pt}
        \vspace{10pt}
        \includegraphics[trim={170 15 25 11},clip,height=4\textwidth]{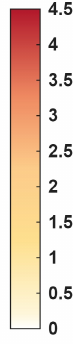}
        \vspace{85pt}
    \end{subfigure}
    \vspace{-40pt} 
    \caption{14-bus system base case generator active and reactive power injection sensitivities.}
    \label{fig:base-case-heatmaps}
    \vspace{-10pt}
\end{figure*}
%
We compute the proposed power flow sensitivity metrics for the Base Case. Fig.~\ref{fig:base-case-heatmaps} maps these metrics and compares them against traditional Impedance Margin Ratio (IMR) values (Fig.~\ref{fig:14busIMR}). While both approaches correctly identify weak IBR buses, the proposed metrics uniquely map this sensitivity directly to specific power flow shifts, rather than relying on conservative, worst-case impedance margins.

\begin{figure}
    \centering
    \includegraphics[width=0.985\linewidth]{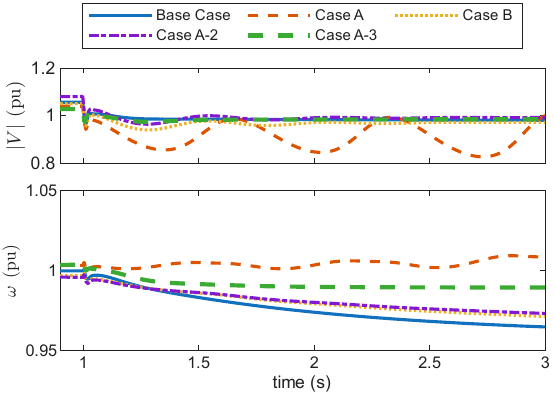}
    \caption{Time-domain response of bus 10 voltage magnitude and frequency for the cases in Table \ref{tab:case-descriptions}. A 0.67 pu load step at bus 5 occurs at $t = 1$ s. Case A exhibits an undamped 1.4 Hz oscillation, which is not prevalent in Case B. Increasing reactive power injected at bus 8 (Case A-2) or changing its control from GFL to GFM (Case A-3) mitigate the oscillation introduced in Case A. 
    }
    \label{fig:load-step-response}
    \vspace{-5pt}
\end{figure}

To validate these indices dynamically, we simulate the nonlinear time-domain system for different cases detailed in Table~\ref{tab:case-descriptions}. We subject the system to an active power load step at bus 5 under two different generator dispatches. While both cases increase generation at bus 10, Case A also increases generation at bus 8 (high PTSI), whereas Case B increases generation at bus 6 (low PTSI). As predicted, the time-domain response in Fig.~\ref{fig:load-step-response} shows that dispatching the highly sensitive bus 8 (Case A) induces a sustained 1.4 Hz oscillation, whereas dispatching the less sensitive bus 6 (Case B) maintains adequate damping.
\setlength{\tabcolsep}{3pt} 
\renewcommand{\arraystretch}{1.2} 
\newcommand{\hquad}{\hspace{0.13em}} 
\begin{table}[!b]
\vspace{-5pt}
    \centering
    \caption{Comparison of predicted and actual eigenvalue perturbations for three modes of the Base Case 14-bus test system.}
    \vspace{-5pt}
    \begin{tabular}{c|c|c|c|c}
        \textbf{Perturbation} & & \thead{$\lambda$ = -10.610 \\ + 0.000\hquad{}j}& \thead{$\lambda$ = -10.791 \\ + 3.382\hquad{}j} & \thead{$\lambda$ = -159.545  \\ + 427.018\hquad{}j} \\
        \cline{3-5}
        $\Delta \mu \times 10^{5}$ & & $\Delta \lambda \times 10^5$ & $\Delta \lambda \times 10^5$ & $\Delta \lambda \times 10^5$ \\
        \hline
         \multirow{3}{*}{$\Delta P_8 = 1$~pu}
         & \scriptsize{Actual} & 0.016 &  0.302 - 0.183\hquad{}j & 0.456 + 0.073\hquad{}j \\
         & \scriptsize{Pred.}  & 0.016 &  0.302 - 0.183\hquad{}j & 0.455 + 0.073\hquad{}j \\
         & \scriptsize{Error}  & 0.28\% & 0.02\% & 0.20\% \\ \hline
         \multirow{3}{*}{$\Delta Q_8 = 1$~pu} 
         & \scriptsize{Actual} & 0.522 & -0.645 + 0.008\hquad{}j & 2.985 + 0.935\hquad{}j \\
         & \scriptsize{Pred.}  & 0.522 & -0.645 + 0.008\hquad{}j & 2.983 + 0.933\hquad{}j \\ 
         & \scriptsize{Error}  & 0.00\% & 0.01\% & 0.09\% \\ \hline
         \multirow{3}{*}{$\Delta V_1 = 1$~pu} 
         & \scriptsize{Actual} & 2.330 & -3.455 - 5.890\hquad{}j & 18.035 + 4.269\hquad{}j \\
         & \scriptsize{Pred.}  & 2.330 & -3.455 - 5.890\hquad{}j & 18.036 + 4.266\hquad{}j \\ 
         & \scriptsize{Error}  & 0.00\% & 0.00\% & 0.02\% \\ \hline
         \multirow{3}{*}{$\Delta B_{5,6} = 1$~pu} 
         & \scriptsize{Actual} & 0.112  & 0.188 - 0.075\hquad{}j  & 20.251 + 1.385\hquad{}j \\
         & \scriptsize{Pred.}  & 0.112  & 0.188 - 0.075\hquad{}j  & 20.248 + 1.386\hquad{}j \\ 
         & \scriptsize{Error}  & 0.04\% & 0.04\% & 0.02\% \\ \hline
    \end{tabular}
    \label{tab:eig-perturbations}
    \vspace{-10pt}
\end{table}
%
%
\begin{figure}
\vspace{0pt}  
    \centering
    \begin{subfigure}[t]{0.20\textwidth}
        \centering
        \includegraphics[trim={15 15 57 15},clip,height=\textwidth]{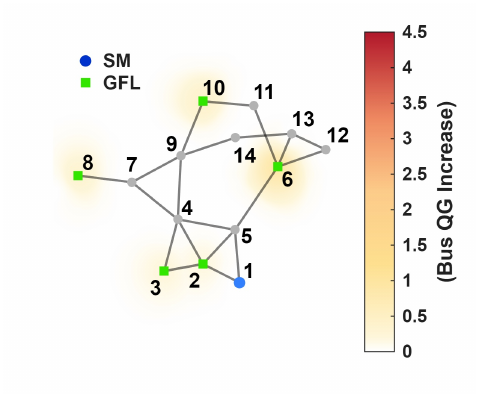}
        \caption{QTSI}
        \label{fig:14busCaseAQtosens}
    \end{subfigure}
    ~
    \begin{subfigure}[t]{0.20\textwidth}
        \centering
        \includegraphics[trim={15 15 57 15},clip,height=\textwidth]{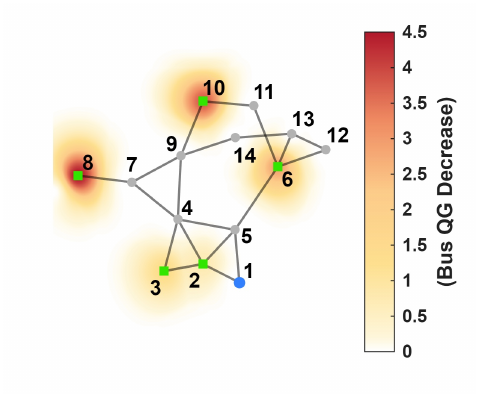}
        \caption{QFSI}
        \label{fig:14busCaseAQfrsens}
    \end{subfigure}
    ~
    \begin{subfigure}[]{0.05\textwidth}
        \centering
        \hspace{-14pt}
        \vspace{10pt}
        \includegraphics[trim={170 15 25 11},clip,height=4\textwidth]{results/14bus/caseA/QGFSI.pdf}
        \vspace{85pt}
    \end{subfigure}
    \vspace{-80pt}
    \caption{14-bus generator reactive power injection sensitivities for case A.}
    \label{fig:case-a-QSI}
\vspace{-10pt}
\end{figure}

Finally, we explore two remedial actions to mitigate the oscillation in Case A. First, guided by the reactive power sensitivities in Fig.~\ref{fig:case-a-QSI}, we observe that bus 8 exhibits a high QFSI (indicating an increase in damping if we increase reactive power injection) and a low QTSI (indicating minimal destabilization risk for this increase in reactive power). Consequently, increasing the reactive power injection at bus 8 (Case A-2) successfully restores modal damping (Fig.~\ref{fig:load-step-response}). 

\begin{figure}
    \centering
    \begin{subfigure}[t]{0.20\textwidth}
        \centering
        \includegraphics[trim={15 15 57 15},clip,height=\textwidth]{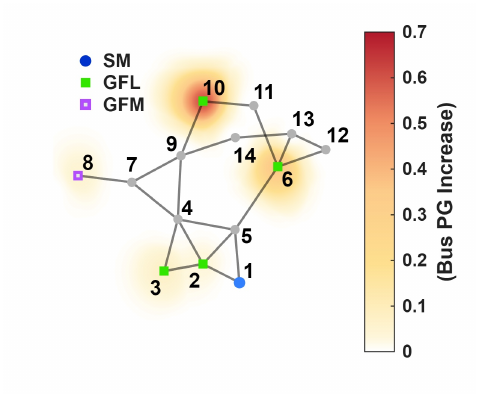}
        \caption{PTSI}
        \label{fig:14busGFMPtosens}
    \end{subfigure}
    ~
    \begin{subfigure}[t]{0.20\textwidth}
        \centering
        \includegraphics[trim={15 15 57 15},clip,height=\textwidth]{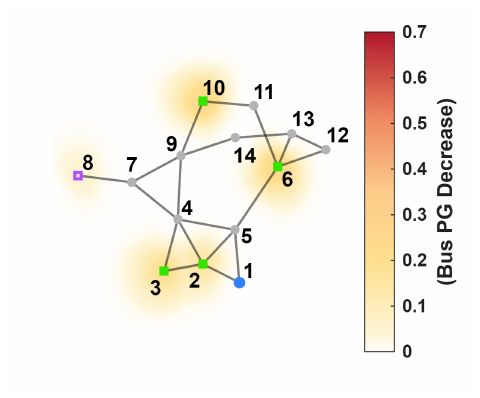}
        \caption{PFSI}
        \label{fig:14busGFMPfrsens}
    \end{subfigure}
    ~
    \begin{subfigure}[]{0.05\textwidth}
        \centering
        \hspace{-14pt}
        \vspace{10pt}
        \includegraphics[trim={170 15 24 11},clip,height=4\textwidth]{results/14bus/GFM/PGFSI.pdf}
        \vspace{85pt}
    \end{subfigure}
    \vspace{-80pt}
    \caption{14-bus generator active power injection sensitivities for base case with GFM IBR control at bus 8.}
    \label{fig:base-case-gfm-Psens}
\vspace{-15pt}
\end{figure}

Alternatively, changing the bus 8 IBR to a grid-forming (GFM) droop controller (Case A-3) fundamentally alters the system's sensitivity. As shown in Fig.~\ref{fig:base-case-gfm-Psens}, the system is no longer sensitive to active power increases at bus 8, which successfully prevents the oscillation during the load step.

%
%
%


\subsection{118-Bus Test System Results}

To evaluate the proposed network admittance and voltage sensitivities, we utilize a modified IEEE 118-bus test system~\cite{christie1993ieee118} comprising a mix of SM and IBR devices (GFLs and GFMs). The SMs are represented by 8th-order models equipped with exciters, power system stabilizers, and governors. SMs and GFMs are modeled as PV buses (with bus 1 serving as the slack bus), while load buses and most GFLs are modeled as PQ buses. The system is initialized with several lines switched off, and loads are modeled as constant impedances linearized around the initial operating point.

\begin{figure*}[t]
    \centering
    \begin{subfigure}[t]{0.5\textwidth}
        \centering
        \includegraphics[trim={65 15 137 11},clip,height=9cm]{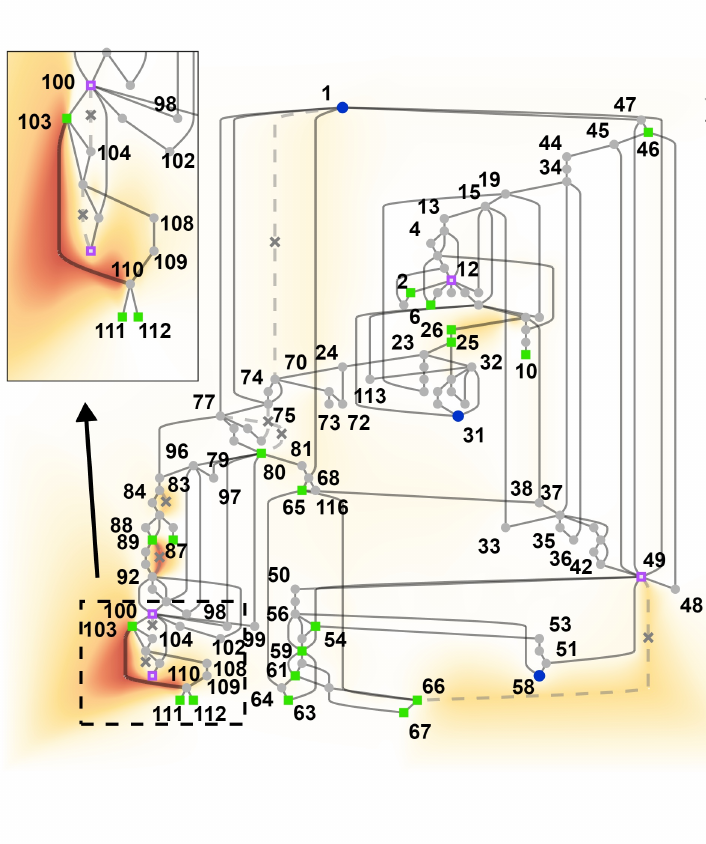}
        
        \vspace{-10pt}
        \caption{$\mathrm{B_{ij}}$FSI}
        \label{fig:118busBijFSI}
    \end{subfigure}%
    \begin{subfigure}[t]{0.5\textwidth}
        \centering
        \hspace{-40pt}
        \includegraphics[trim={65 15 65 11},clip,height=9cm]{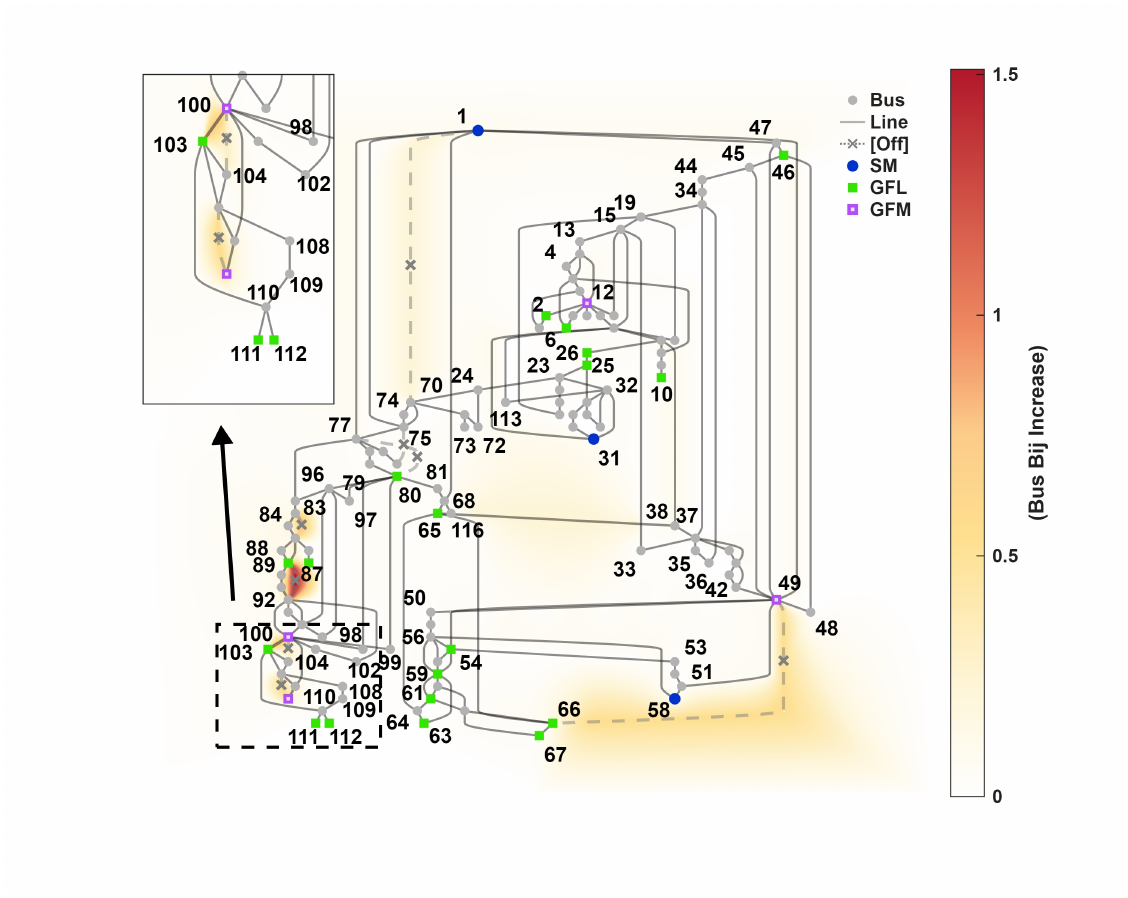}
        
        \vspace{-10pt}
        \caption{$\mathrm{B_{ij}}$TSI}
        \label{fig:118busBijTSI}
    \end{subfigure}%
    \begin{subfigure}[]{0.09\textwidth}
        \centering
        \hspace{-65pt}
        \vspace{250pt}
        \includegraphics[trim={475 15 40 11},clip,height=9cm]{results/118bus/BijFSI_GFL-Line_Inlay.pdf}
    \end{subfigure}
    \vspace{-250pt}
    
    \caption{118-bus line admittance sensitivities.
    }
    \label{fig:118bus-line-sens}
\vspace{-10pt}
\end{figure*}

We calculate the line admittance sensitivity indices for a decrease ($\mathrm{B_{ij}}$FSI) and an increase ($\mathrm{B_{ij}}$TSI) in line admittance, plotted in Fig.~\ref{fig:118busBijFSI} and Fig.~\ref{fig:118busBijTSI}, respectively. 
Fig.\ref{fig:118busBijFSI} shows that the system exhibits a high $\mathrm{B_{ij}}$FSI for line 103-110, indicating severe sensitivity to its outage, while line 100-103 shows very low sensitivity. Conversely, the $\mathrm{B_{ij}}$TSI metric is also valuable for assessing grid reconnection; Fig.~\ref{fig:118busBijTSI} shows that line 89-92, which is initially open, has the highest sensitivity to admittance increases indicating that switching this line on could unexpectedly degrade small-signal stability. Similarly, the medium $\mathrm{B_{ij}}$TSI sensitivity of line 100-103 indicates that switching this line off can improve the small-signal stability. 

\begin{figure}
    \centering
    \begin{subfigure}[t]{0.45\textwidth}
        \centering
        \includegraphics[width=0.985\linewidth, trim={0 0 0 0},clip]{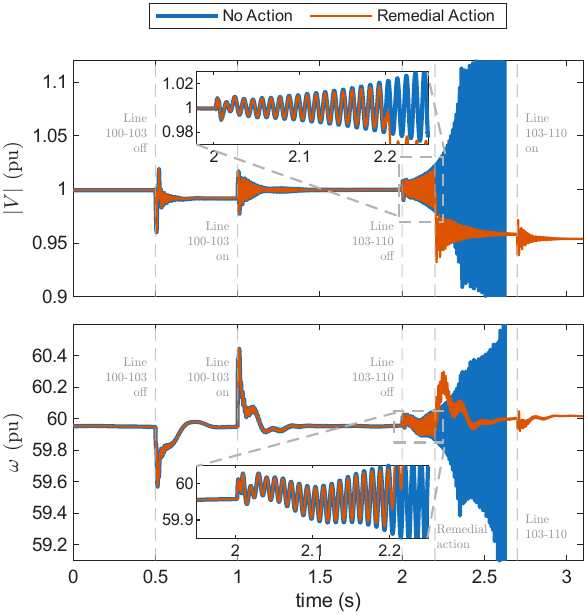}
        \caption{Bus 103 GFL}
        \label{fig:118busLineOff103}
    \end{subfigure}
    
    \begin{subfigure}[t]{0.45\textwidth}
        \centering
        \includegraphics[width=0.985\linewidth, trim={0 0 0 20},clip]{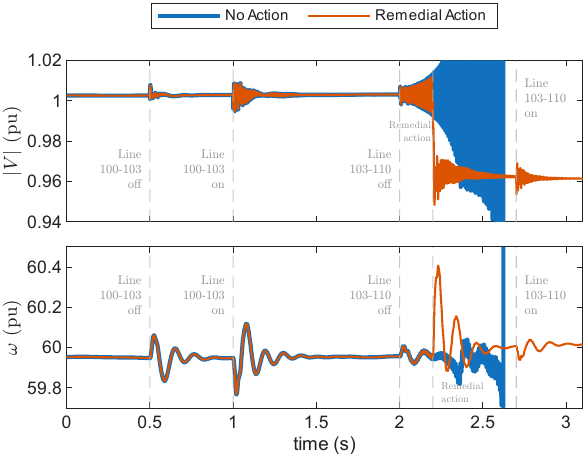}
        \caption{Bus 100 GFM}
        \label{fig:118busLineOff101}
    \end{subfigure}%
    \caption{Time-domain response of bus 103 and bus 100 voltage magnitude and measured frequency when switching the line between bus 100 and bus 103 off at time $t = 0.5$~s and on at $t = 1.0$~s, then switching line 103-110 off at $t = 0.2$~s and on at $t = 2.7$~s. The Bus 103 and bus 100 frequency measurements from their respective GFL PLL and the GFM droop controller. Following the line 103-110 switch, we see a growing undamped $95$~Hz oscillation. The system becomes unstable if no action is taken; we do not plot this unstable 'no action' trace past $t\approx2.6$~s. Applying a remedial action decreasing the voltage at bus 100 by $0.1$~p.u and decreasing $Q_G$ at bus 112 by $0.3$~p.u at $t=2.2$~s resolves the instability. This oscillation occurs due to a control interaction between the GFL's inner control loops and the lines in the area. }
    \label{fig:118busLineSwitchSim}
\vspace{-15pt}
\end{figure}

To validate these analytical predictions, Fig.~\ref{fig:118busLineSwitchSim} illustrates the time-domain response to switching events. 
When line 100-103 is temporarily tripped at $t = 0.5~\si{\second}$ and restored at $t = 1.0~\si{\second}$, the induced oscillations are quickly damped, and the system reaches a new stable steady-state. However, when the highly sensitive line 103-110 is tripped at $t = 2~\si{\second}$, the system immediately becomes unstable, exhibiting a growing, undamped $95~\si{\hertz}$ oscillation that diverges if no action is taken.

\begin{figure}
    \centering
    \begin{subfigure}[t]{0.22\textwidth}
        \centering
        \includegraphics[trim={65 232 380 25},clip,height=\textwidth]{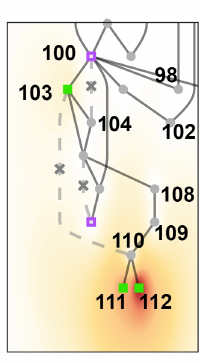}
        \caption{QTSI}
        \label{fig:118busLineOffQtosens}
    \end{subfigure}\hspace{-1.5em}%
    \begin{subfigure}[]{0.053\textwidth}
        \centering
        \vspace{5pt}
        \includegraphics[trim={165 15 25 11},clip,height=4\textwidth]{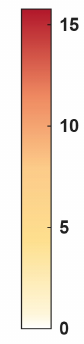}
        \vspace{100pt}
    \end{subfigure}\hspace{-0.75em}%
    \begin{subfigure}[t]{0.22\textwidth}
        \centering
        \includegraphics[trim={65 232 375 25},clip,height=\textwidth]{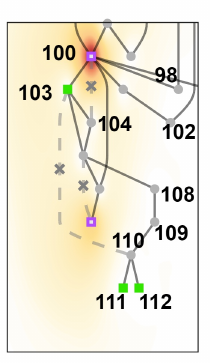}
        \caption{VTSI}
        \label{fig:118busLineOffVtosens}
    \end{subfigure}\hspace{-1.3em}%
    \begin{subfigure}[]{0.053\textwidth}
        \centering
        \vspace{5pt}
        \includegraphics[trim={163 15 25 11},clip,height=4\textwidth]{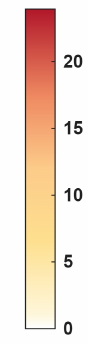}
        \vspace{100pt}
    \end{subfigure}
    \vspace{-95pt}
    \caption{Generator reactive power injection and voltage increase unstable eigenvalue ($\lambda = 0.25+95.63j~\si{\hertz}$) sensitivities for a portion of the 118-bus system with line 103-110 switched off.}
    \label{fig:118busLineOffRemedial}
\vspace{-15pt}
\end{figure}

To identify  remedial actions for the N-1 contingency of line 103-110, we compute the reactive power and voltage magnitude sensitivities specifically for the unstable mode ($\lambda = 0.25+95.63j~\si{\hertz}$). As shown in Fig.~\ref{fig:118busLineOffRemedial}, the unstable mode is highly sensitive to increases in the reactive power at bus 112 and the voltage at bus 100. By applying remedial actions at $t = 2.2~\si{\second}$ (decreasing both $Q_{112}$ and $V_{100}$) the $95~\si{\hertz}$ oscillation is successfully damped, stabilizing the system before the line is restored at $t = 2.7~\si{\second}$.

While first-order predictions of the eigenvalue sensitivities may not provide perfect predictions of the nonlinear trace of the eigenvalues for large steps in a parameter, such as when switching a line admittance off, this case study demonstrates that the proposed indices can provide valuable insights into which lines require further detailed studies to understand their impact on system stability.
%
%

\section{Conclusion} 
\label{sec:conclusion}
%
This paper presents a method to compute power flow sensitivities from perturbations in power flow parameters (power injections and voltages) and network admittances. We use these sensitivities to determine the sensitivities of an AC power system's eigenvalues. We define grid strength metrics based on these eigenvalue sensitivities, and demonstrate the capabilities of these metrics with 14-bus and 118-bus test systems. These metrics can be used to screen dispatches and identify remedial actions for small-signal stability issues. Future work includes using these sensitivities to improve the robustness of the network's small-signal stability, considering how frequency and voltage droop parameters impact resulting power flow perturbations, and utilizing second-order eigenvalue sensitivities to more accurately predict shifts in the eigenvalues. 


\appendix
\subsection{Proof of Device Impedance Sensitivity}
\label{app:proof}

The closed-loop whole-system admittance matrix $\hat{\mathbf{Y}}(s)$ is defined as the relationship between the grid network admittance $\mathbf{Y}_\mathrm{net}(s)$ and the aggregated device impedance $\mathbf{Z}_\mathrm{dev}(s)$:
\begin{equation}
    \hat{\mathbf{Y}} = (I + \mathbf{Y}_\mathrm{net}\mathbf{Z}_\mathrm{dev})^{-1} \mathbf{Y}_\mathrm{net}
\end{equation}
To find the first-order sensitivity of the system with respect to a local power flow parameter $x$ that only affects the device control state, we first invert the whole-system admittance matrix to find the whole-system impedance $\hat{\mathbf{Z}}$:
\begin{equation}
    \hat{\mathbf{Y}}^{-1} = \mathbf{Y}_\mathrm{net}^{-1} (I + \mathbf{Y}_\mathrm{net}\mathbf{Z}_\mathrm{dev})
\end{equation}
Distributing the inverse network admittance yields:
\begin{equation}
    \hat{\mathbf{Y}}^{-1} = \mathbf{Y}_\mathrm{net}^{-1} + \mathbf{Y}_\mathrm{net}^{-1}\mathbf{Y}_\mathrm{net}\mathbf{Z}_\mathrm{dev} = \mathbf{Y}_\mathrm{net}^{-1} + \mathbf{Z}_\mathrm{dev}
\end{equation}
Taking the partial derivative of both sides with respect to the power flow parameter $x$:
\begin{equation}
    \frac{\partial \hat{\mathbf{Y}}^{-1}}{\partial x} = \frac{\partial \mathbf{Y}_\mathrm{net}^{-1}}{\partial x} + \frac{\partial \mathbf{Z}_\mathrm{dev}}{\partial x}
\end{equation}
Because the physical network topology $\mathbf{Y}_\mathrm{net}$ is independent of the device power flow setpoints, its partial derivative with respect to $x$ is zero. Therefore, the sensitivity of the inverse whole-system admittance reduces exactly to the sensitivity of the device impedance:
\begin{equation}
    \frac{\partial \hat{\mathbf{Y}}^{-1}}{\partial x} = \frac{\partial \mathbf{Z}_\mathrm{dev}}{\partial x}
\end{equation}

\subsection{Proof of Network Admittance Sensitivity}
\label{app:netproof}

Following the derivation in Appendix~\ref{app:proof}, the inverse of the whole-system admittance matrix is analytically defined as the sum of the inverse network admittance and the device impedance:
\begin{equation}
    \hat{\mathbf{Y}}^{-1} = \mathbf{Y}_\mathrm{net}^{-1} + \mathbf{Z}_\mathrm{dev}
\end{equation}
To find the sensitivity of the system with respect to a change in the physical network topology, we take the partial derivative of both sides with respect to a specific line susceptance parameter $B_{ij}$:
\begin{equation}
    \frac{\partial \hat{\mathbf{Y}}^{-1}}{\partial B_{ij}} = \frac{\partial \mathbf{Y}_\mathrm{net}^{-1}}{\partial B_{ij}} + \frac{\partial \mathbf{Z}_\mathrm{dev}}{\partial B_{ij}}
\end{equation}

Applying the standard matrix derivative identity for an inverse matrix ($\frac{\partial \mathbf{A}^{-1}}{\partial x} = -\mathbf{A}^{-1} \frac{\partial \mathbf{A}}{\partial x} \mathbf{A}^{-1}$), the sensitivity of the inverse whole-system admittance evaluates to:
\begin{equation}
    \frac{\partial \hat{\mathbf{Y}}^{-1}}{\partial B_{ij}} = -\mathbf{Y}_\mathrm{net}^{-1} \frac{\partial \mathbf{Y}_\mathrm{net}}{\partial B_{ij}} \mathbf{Y}_\mathrm{net}^{-1}+ \frac{\partial \mathbf{Z}_\mathrm{dev}}{\partial B_{ij}}.
\end{equation}

\subsection{Proposed Sensitivity Evaluation Procedure}
\label{sec:evaluation_procedure}

To evaluate the small-signal stability impact of an arbitrary grid parameter $x$ (representing either a structural network element or a nodal operating setpoint) the following procedure can be applied:
\vspace{1em} 

\noindent \textbf{Inputs:}
\begin{itemize}
    \item Baseline steady-state power flow solution, topological line data, and the selected perturbation parameter ($x$).
    \item Baseline localized device equivalent impedance matrices ($\mathbf{Z}_\mathrm{dev}$). These can be formulated analytically from white-box models or sampled from black-box models using frequency scanning at the baseline operating point.
    \item Device equivalent impedance sensitivities with respect to local power flow parameters ($\frac{\partial \mathbf{Z}_\mathrm{dev}}{\partial \mathbf{PF}}$), obtained analytically or via numerical perturbation.
\end{itemize}

\noindent \textbf{Output:}
\begin{itemize}
    \item The first-order sensitivity of the selected eigenvalue with respect to the grid parameter ($\frac{\partial \lambda}{\partial x}$).
\end{itemize}

\vspace{1em} 

\noindent \textbf{Step 1: Baseline System Initialization.} Solve the baseline steady-state power flow for the unperturbed system. Using this operating point, formulate the baseline network admittance matrix $\mathbf{Y}_\mathrm{net}$ and the device impedance matrix $\mathbf{Z}_\mathrm{dev}$ with the device equivalent impedances oriented to the global reference frame. Construct the whole system admittance matrix $\hat{\mathbf{Y}}$, as \eqref{eq:whole-system-admittance}, and solve for the system's poles/eigenvalues. Identify the mode of interest $\lambda$ and extract the associated residue matrix $\mathrm{Res}_\lambda \hat{\mathbf{Y}}$ from the whole system admittance matrix.

\noindent \textbf{Step 2: Network Structural Sensitivity.} Evaluate the direct topological impact of the parameter $x$ on the physical transmission network by calculating $\frac{\partial \mathbf{Y}_\mathrm{net}}{\partial x}$. If $x$ represents a line admittance parameter, this is derived as \eqref{eq:network-line-sensitivity}. If $x$ represents a nodal power injection, the physical network topology is unaltered, and this term evaluates strictly to zero.

\noindent \textbf{Step 3: Steady-State Power Flow Sensitivity.} Determine the global steady-state power flow shifts induced by the parameter $x$. Using the partitioned inverse power flow Jacobian, calculate the sensitivities of the dependent bus voltage magnitudes and angles across the network and the dependent bus power injections as described in Section \ref{sec:power-flow-sensitivity} to maintain the required power balances.

\noindent \textbf{Step 4: Device Impedance Sensitivity.} Map the power flow shifts from Step 3 to the equivalent admittance models of the individual devices. Using the chain rule, calculate the total device impedance sensitivity $\frac{\partial \mathbf{Z}_\mathrm{dev}}{\partial x}$ by scaling the device-specific power flow sensitivities by the parameter $x$ specific power flow shifts.

\noindent \textbf{Step 5: Total Eigenvalue Sensitivity.} Consolidate the structural and device-level pathways to find the total eigenvalue sensitivity. Using the linearity of the inner product, calculate the final sensitivity by projecting both the network and device variations onto the baseline residue:
\begin{equation}
    \frac{\partial \lambda}{\partial x} = \langle \mathrm{Res}_{\lambda}^* \hat{\mathbf{Y}}, \mathbf{Y}_\mathrm{net}^{-1} \frac{\partial \mathbf{Y}_\mathrm{net}}{\partial x} \mathbf{Y}_\mathrm{net}^{-1} \rangle + \sum_{k \in \mathcal{N}} \langle -\mathrm{Res}_{\lambda}^* \hat{\mathbf{Y}}_{kk}, \frac{\partial Z_k}{\partial x} \rangle
\end{equation}

\textit{Note:} To maximize computational efficiency during repeated evaluations across large networks, we can save intermediate terms that are dependent only on the base values and the eigenvalue, such as the residue of the whole system admittance matrix. The projection of the baseline matrices used for calculating the sensitivity to the structural perturbation in Step 5 can also be pre-computed and stored. By defining a sparse sensitivity projection matrix $\mathbf{W}_\mathrm{net}(\lambda) = \mathbf{Y}_\mathrm{net}^{-1}(\lambda) (\mathrm{Res}_{\lambda}^* \hat{\mathbf{Y}}) \mathbf{Y}_\mathrm{net}^{-1}(\lambda)$, we avoid inverting the network admittance matrices repeatedly. The eigenvalue sensitivity from the structural perturbation can then be calculated as $\langle \mathbf{W}_\mathrm{net}, \frac{\partial \mathbf{Y}_\mathrm{net}}{\partial x} \rangle$.

\bibliographystyle{IEEEtran}
\bibliography{Reference}

\newpage

 



\vfill

\end{document}